\input{aipcheck}

\documentclass[
    ,final            % use final for the camera ready runs
  ]
  {aipproc}

\layoutstyle{6x9}

\begin{document}

\title{COSY-11: an experimental facility for studying meson production 
in free and quasi-free nucleon-nucleon collisions.}

\classification{13.85.Ni}
\keywords      {meson production, experimental facillity}

\author{P. Klaja}{
  address={Nuclear Physics Department, Jagellonian University, 30-059 Cracow}
}

\author{H.-H.~Adam}{
  address={Institut f{\"u}r Kernphysik, Universist\"at M\"unster, 48419 M\"unster, Germany}
}

\author{A.~Budzanowski}{
  address={Institute of Nuclear Physics, 31-342 Cracow, Poland}
}

\author{R.Czy{\.z}ykiewicz}{
   address={Institut f\"ur Kernpfysik, Forschungszentrum J\"ulich, 52425 J\"ulich, Germany}
  ,altaddress={Nuclear Physics Department, Jagellonian University, 30-059 Cracow}
}

\author{D.~Grzonka}{
   address={Institut f\"ur Kernpfysik, Forschungszentrum J\"ulich, 52425 J\"ulich, Germany}
}

\author{M.~Janusz}{
  address={Nuclear Physics Department, Jagellonian University, 30-059 Cracow}
}

\author{L.~Jarczyk}{
  address={Nuclear Physics Department, Jagellonian University, 30-059 Cracow}
}

\author{B.~Kamys}{
  address={Nuclear Physics Department, Jagellonian University, 30-059 Cracow}
}

\author{A.~Khoukaz}{
  address={Institut f{\"u}r Kernphysik, Universist\"at M\"unster, 48419 M\"unster, Germany}
}

\author{K.~Kilian}{
   address={Institut f\"ur Kernpfysik, Forschungszentrum J\"ulich, 52425 J\"ulich, Germany}
}

\author{P.~Moskal}{
  address={Nuclear Physics Department, Jagellonian University, 30-059 Cracow}
  ,altaddress={Institut f\"ur Kernpfysik, Forschungszentrum J\"ulich, 52425 J\"ulich, Germany}
}

\author{W.~Oelert}{
   address={Institut f\"ur Kernpfysik, Forschungszentrum J\"ulich, 52425 J\"ulich, Germany}
}

\author{C.~Piskor--Ignatowicz}{
  address={Nuclear Physics Department, Jagellonian University, 30-059 Cracow}
}

\author{J.~Przerwa}{
  address={Nuclear Physics Department, Jagellonian University, 30-059 Cracow}
}

\author{J.~Ritman}{
  address={Institut f\"ur Kernpfysik, Forschungszentrum J\"ulich, 52425 J\"ulich, Germany}
}

\author{T.~Ro\.zek}{
    address={Institut f\"ur Kernpfysik, Forschungszentrum J\"ulich, 52425 J\"ulich, Germany}
   ,altaddress={Institute of Physics, University of Silesia, 40-007 Katowice, Poland}
}

\author{T.~Sefzick}{
  address={Institut f\"ur Kernpfysik, Forschungszentrum J\"ulich, 52425 J\"ulich, Germany}
}

\author{M.~Siemaszko}{
  address={Institute of Physics, University of Silesia, 40-007 Katowice, Poland}
}

\author{J.~Smyrski}{
  address={Nuclear Physics Department, Jagellonian University, 30-059 Cracow}
}

\author{A.T\"aschner}{
   address={Institut f{\"u}r Kernphysik, Universist\"at M\"unster, 48419 M\"unster, Germany}
}

\author{J.~Wessels}{
    address={Institut f{\"u}r Kernphysik, Universist\"at M\"unster, 48419 M\"unster, Germany}
}

\author{P.~Winter}{
  address={Institut f\"ur Kernpfysik, Forschungszentrum J\"ulich, 52425 J\"ulich, Germany}
}

\author{M.~Wolke}{
  address={Institut f\"ur Kernpfysik, Forschungszentrum J\"ulich, 52425 J\"ulich, Germany}
}

\author{P.~W\"ustner}{
  address={ZEL Forschungszentrum J\"ulich, 52425 J\"ulich, Germany}
}
\author{W.~Zipper}{
  address={Institute of Physics, University of Silesia, 40-007 Katowice, Poland}
}

\begin{abstract}
The COSY-11 experimental setup  is an internal facility installed at 
the COoler SYnchrotron COSY in J{\"u}lich. It allows to investigate 
meson production in free and quasi-free nucleon-nucleon collisions, 
eg. $pp \rightarrow pp meson$ and $pd \rightarrow p_{sp}np meson$ 
reactions. 
Drift chambers and scintillators permit to measure outgoing protons, 
separated in magnetic field of COSY-11 dipole.
Neutrons are registered in the neutron modular detector installed 
downstream the beam.
Recently, the experimental setup has been extended with spectator 
detector, deuteron drift chamber  and polarization monitoring system, 
and since then meson production can be investigated also as a function 
of spin and isospin of colliding nucleons.\\
%The COSY-11 facility, as well as experimental methods developed to 
%identify events corresponding to the meson production in free and 
%quasi-free nucleon-nucleon collisions will be presented.

% This template file shows how to use the \texttt{aipproc} class to
% produce a paper with the correct layout for \emph{%
%   AIP Conference Proceedings  6in   x 9in single column}.

% A full description of the features supported by the \texttt{aipproc}
% class can be found in the \texttt{aipguide.pdf} document accompanying
% the distribution.

% Frequently asked questions can be found in the \texttt{FAQ.txt}
% document.
\end{abstract}

\maketitle

%%%%%%%%%%%%%%%%%%%%%%%%%%%%%%%%%%%%%%%%%%%%
%% MAINMATTER
%%%%%%%%%%%%%%%%%%%%%%%%%%%%%%%%%%%%%%%%%%%%

\section{COSY-11 experimental setup}

%%%%%%%\subsection{Overview}

The COSY-11 \cite{brauksiepe, pawel, smyrski} is an internal 
facility at the 
COoler SYnchrotron COSY \cite{prashun, maier} in J{\"u}lich, Germany. 
The recently extended experimental setup is presented in figure 1. It 
permits to investigate meson and hyperon production in proton-proton 
and proton-deuteron reactions close to the kinematical threshold. 
A very high experimental precision achieved for the four-momenta 
determination allows to study the excitation function for the hyperons 
\cite{kowina, sewerin} and mesons \cite{smyrski1} creation down to 
the fraction of 1 MeV above the threshold.
In general the reactions are studied by the determination of the 
four momentum vectors of colliding and outgoing nucleons (in case of 
the $pp \rightarrow pYK^{+}$ also four momentum vector of the 
$K^{+}$ meson is  determined).
\begin{figure}
  \includegraphics[height=.32\textheight]{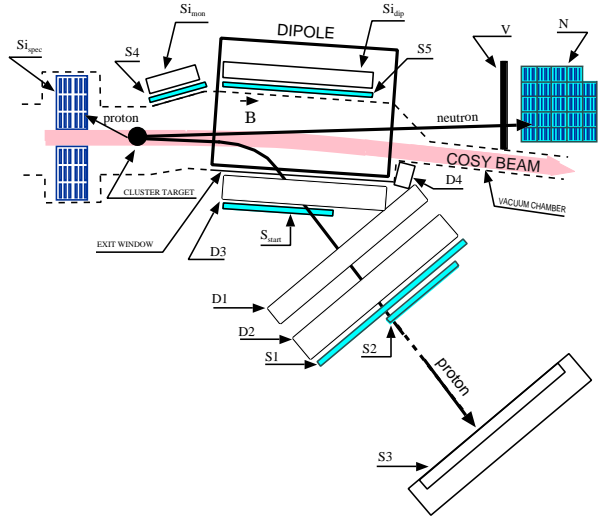}
  \caption{Schematic view of COSY-11 detection setup \cite{brauksiepe}.  
           D1, D2, D3 and D4 denote the drift chambers; 
           S1, S2, S3, S4, S5, $S_{start}$ and V 
           the scintillation detectors; N the neutron detector and 
            $Si_{mon}$, $Si_{spec}$ and $Si_{dip}$ silicon strip 
             detectors 
             to detect elastically scattered, spectator protons 
             and negatively charged particles, 
             respectively.}
\end{figure}
In front of a normal C-type COSY bending magnet, there is installed 
a cluster target. Both, hydrogen or deuteron targets \cite{dombrowski, 
khoukaz} can be used.\\
Due to smaller momenta of positively charged reaction products, they 
are separated from the circulating beam in the magnetic field of the 
COSY magnet and diverted towards the detection setup. Passing through 
a specially developed large exit foil of a vacuum chamber, which is 
mounted inside the dipole gap, the ejectiles reach the detection 
system operated in normal atmosphere \cite{brauksiepe}.\\
Negatively charged particles, with their tracks being bent towards 
the inside of the dipole gap, are detected by an array of silicon 
detectors ($Si_{dip}$) and an additional scintilation counter (S5), 
both mounted inside 
the gap \cite{brauksiepe}.\\
Neutral reaction ejectiles are detected with the modular neutral 
particles detector (N) installed downstream the beam \cite{przerwa}.\\
The detection setup has been extended with spectator detector 
($Si_{spec}$) \cite{biliger} 
and beam polarization monitoring system \cite{rafal, czyzyk}. Therefore 
at present, the meson production can be 
investigated also as a function of spin \cite{winter} and isospin 
\cite{pmoskal} of colliding 
nucleons. The newly installed deuteron drift chamber (D4) \cite{cezary} 
permits to study also meson production in $pd \rightarrow pdX$ 
reactions. The spectator detector is installed inside 
the vacuum chamber as it is shown in the figure 1. Part of the COSY-11 
detection setup used for the registration of elastically scattered 
protons in the proton-proton free and quasi-free collisions, is shown in 
figure 1. Trajectories of protons scattered in the forward direction 
are measured by means of two drift chambers \cite{smyrski} 
and a scintillation hodoscope (S1), whereas the recoil protons 
are registered in coincidence with forward ones using a silicon pad 
detector arrangement ($Si_{mon}$) and scintillation detector (S4) 
\cite{pawel}.  

\section{Experimental methods}

%\subsection{<A subsection>}

Identification of registered particles proceeds including various 
methods. Positively charged ejectiles are identified by independent 
measurements of their momentum and velocity. For neutrons and gamma 
quanta velocity vector is obtained. Spectator protons are identified 
by measurements of their kinetic energy and momentum direction.\\
For the $pp \rightarrow ppX$ reaction, the collision of protons 
results in the production of meson. Two ejected protons have smaller 
momenta and they are separated in the dipole magnetic field from the 
circulating beam. For this type of reactions, 
the hardware trigger \cite{magnus}, based on signals from scintillation 
detectors, was 
adjusted to register all events with at least two positively 
charged particles. Tracking back trajectories from drift chambers 
\cite{smyrski} 
through the dipole magnetic field to the target point allowed for the 
determination of the particles momenta. Having momentum and velocity, 
the latter measured using scintillation detectors, it is possible 
to identify the mass of the particle \cite{hab}. As an example, figure 2 
shows the squared mass of two simultaneously detected particles for 
$pp \rightarrow A^{+}B^{+}X$ reaction (left panel) and 
$dp \rightarrow A^{+}B^{+}X$ reaction (right panel).     
Measured reactions can be grouped according to the type of 
ejectiles. The reaction with 
two protons, proton and pion, proton and deuteron, pion and deuteron 
and two pions can be very clearly separeted.

\begin{figure}[h]
  \includegraphics[height=.21\textheight]{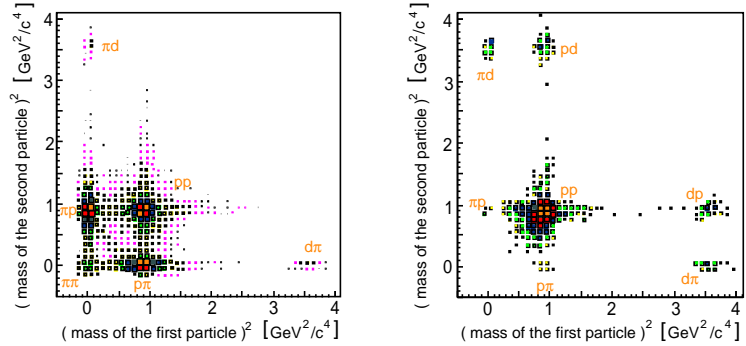}
  \caption{{\bf Left:} Squared masses of two positively charged particles 
                   measured in coincidence in the 
                   $pp \rightarrow A^{+}B^{+}X$ 
                   reaction \cite{pawel1}.
          {\bf Right:} Squared masses of two positively charged particles 
                   measured in coincidence in the
                   $dp \rightarrow A^{+}B^{+}X$ 
                   reaction \cite{joanna}.}
\end{figure}
The knowledge of the momenta of both protons, for example in the 
$pp \rightarrow ppX$ reaction, before and after the reaction allows 
to calculate the mass of an unobserved particle or system of particles 
created in the reaction. An unobserved meson is identified 
via the missing mass technique.
Figure 3 demonstrates the achieved missing mass resolution for the 
$pp \rightarrow ppX$ reaction measured close to the $\eta'$ meson 
production threshold, at the COSY-11 detection system, when using a 
stochastically 
cooled beam \cite{hab}. It is worth noting that the experimental mass resolution is comparable with the natural width of the $\eta'$ meson 
($\Gamma_{\eta'} = 0.202$ MeV \cite{hagiwara}). 
\begin{figure}
  \includegraphics[height=.2\textheight]{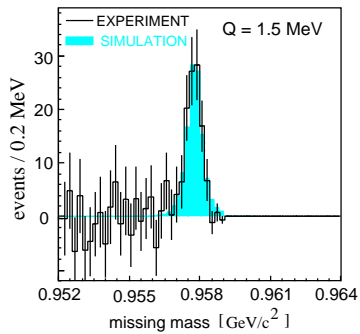}
  \caption{ Missing mass distribution for the $pp \rightarrow ppX$ 
            reaction measured close to the $\eta'$ meson production 
            threshold \cite{pawel2}}
\end{figure}
\begin{theacknowledgments}
The work has been supported by the
European Community - Access to
Research Infrastructure action of the
Improving Human Potential Programme,
by the FFE grants (41266606 and 41266654) from the Research Centre J{\"u}lich,
by the DAAD Exchange Programme (PPP-Polen),
by the Polish State Committe for Scientific Research
(grant No. PB1060/P03/2004/26),
and by the RII3/CT/2004/506078
- Hadron Physics-Activity -N4:EtaMesonNet.

\end{theacknowledgments}



\bibliographystyle{aipproc}   % if natbib is available

%%%\bibliography{sample}

%%%\IfFileExists{\jobname.bbl}{}
%%% {\typeout{}
%%%  \typeout{******************************************}
%%%  \typeout{** Please run "bibtex \jobname" to optain}
%%%  \typeout{** the bibliography and then re-run LaTeX}
%%%  \typeout{** twice to fix the references!}
%%%  \typeout{******************************************}
%%%  \typeout{}
%%% }

%%%\end{document}

%%%%%%%%%%%%%%%%%%%%%%%%%%%%%%%%%%%%%%%%%%%
%% The following lines show an example how to produce a bibliography
%% without the help of the BibTeX program. This could be used instead
%% of the above.
%%%%%%%%%%%%%%%%%%%%%%%%%%%%%%%%%%%%%%%%%%%

\end{document}